\documentclass[a4paper, oneside, twocolumn, notitlepage, 10pt]{extarticle_ecoc}
\usepackage{ecoc}
\usepackage{cleveref} 
\crefname{figure}{fig.}{figs}           
\usepackage{biblatex}

\begin{document}
\selectlanguage{english}    


\title{1.6 Tbps Classical Channel Coexistence With DV-QKD Over Hollow Core Nested Antiresonant Nodeless Fibre (HC-NANF)}%


\author{
    O.~Alia\textsuperscript{(1)}*, R.~S.~Tessinari\textsuperscript{(1)},
    T.~D.~Bradley\textsuperscript{(2)},
    H.~Sakr\textsuperscript{(2)},
    K.~Harrington\textsuperscript{(2)},
    J.~Hayes\textsuperscript{(2)},
    Y.~Chen\textsuperscript{(2)},\\
    P.~Petropoulos\textsuperscript{(2)},
    D.~Richardson\textsuperscript{(2)},
    F.~Poletti\textsuperscript{(2)}
    G.~T.~Kanellos\textsuperscript{(1)},
    R.~Nejabati\textsuperscript{(1)},
    D.~Simeonidou\textsuperscript{(1)}}

\maketitle                  


\begin{strip}
 \begin{author_descr}
    \centering
    
   \textsuperscript{(1)} High Performance Networks Group, University of Bristol, Woodland Road, Bristol, UK\\
   \textsuperscript{(2)} Optoelectronics Research Centre, University of Southampton, Southampton, SO17 1BJ, UK\\
   {{*\textit{Corresponding author}: \textit{obada.alia@bristol.ac.uk}}}

 \end{author_descr}
\end{strip}

\setstretch{1.1}
\renewcommand\footnotemark{}
\renewcommand\footnoterule{}
\let\thefootnote\relax\footnotetext{978-1-6654-3868-1/21/\$31.00 \textcopyright 2021 IEEE}


\begin{strip}
  \begin{ecoc_abstract}
    
    We demonstrate for the first time the coexistence of a quantum-channel and 8$\times$200 Gpbs 16-QAM optical channels with launching powers as high as -9dBm/channel in a 2~km HC-NANF. Comparative analysis with single-mode fibre reveals that the quantum-channel could not be sustained at such power-levels.

  \end{ecoc_abstract}
\end{strip}


\section{Introduction}
Quantum Key Distribution (QKD) technology has been considered as the ultimate physical layer security due to its dependencies on the physical laws of physics to generate quantum keys \cite{pirandola2020advances}. However, in order for QKD to become functional for practical scenarios, it must be integrated with the classical optical networking infrastructure. Coexisting quantum and classical channels represent a challenge for QKD to multiple factors such as the high optical power used on classical channels (orders of magnitude higher than for quantum communication) and the additional noise generated from such channels due to optical non-linear effects \cite{chraplyvy1990limitations} or insufficient isolation between the classical and quantum channels. This additional noise degrades the quantum channel performance and was studied extensively \cite{eraerds2010quantum, patel2012coexistence, zavitsanos2019coexistence,da2014impact, subacius2005backscattering}.

Coping with optical nonlinearity represents a major challenge for QKD systems and sharp filters are required to partly isolate the noise from the classical channels and improve the quantum signal. However, non-linear processes in glass fibres also generate excessive photons at the same wavelength as the quantum channel that cannot be mitigated, dramatically affecting its performance. HCFs offer a radical solution as they provide several attractive advantages compared to glass core fibres, allowing for an ultra-low optical mode overlap with the glass (and hence reduced optical nonlinearity and Rayleigh scattering), a lower latency, and a very low total chromatic dispersion. These desirable qualities enable the transmission of classical channels at high optical powers \cite{poletti2013towards, liu2019nonlinearity, russell2014hollow} while coexisting with quantum channels over the same medium . One associated problem of HCF is the increased fibre losses and the high loss of interfacing between HCF and single mode fibre (SMF). A novel HCF design that could reach a total loss value lower than that of conventional solid fibres was proposed in 2014 and is called Hollow Core Nested Antiresonant Nodeless Fibre (HC-NANF) \cite{poletti2014nested}. The losses of HC-NANF have improved immensely from 1.3~dB/km \cite{bradley2018record} to 0.65~dB/km \cite{bradley2019antiresonant} and recently to 0.28~dB/km \cite{jasion2020hollow}. Moreover, a record-low loss of interconnection between HC-NANF and SMF of 0.15~dB has been achieved which is slightly higher than the theoretically-expected minimum loss of 0.08~dB \cite{suslov2021low}. HC-NANF was also used in long-haul WDM transmission with a record transmission distance of 618~km and 201~km using PM-QPSK and 16-QAM respectively. \cite{nespola2020transmission}.


In this paper, we take the advantage of the ultra-low nonlinearity of HC-NANF to demonstrate the coexistence of a 16~dB power budget discrete variable QKD (DV-QKD) technology \cite{CL3} and 8 $\times$ 200~Gbps 16-QAM carrier-grade classical optical channels at an extremely high coexistence power of 0~dBm over a 2~km HC-NANF, revealing minimal effects on the quantum channel performance. We also compare the QKD performance using SMF and HC-NANF in terms of Secret Key Rate (SKR) and Quantum Bit Error Rate (QBER) to prove the superiority of HC-NANF for the coexistence of quantum and classical channels.

\section{Testbed}

\begin{figure*}[t]
   \centering
        \includegraphics[width=0.9\linewidth]{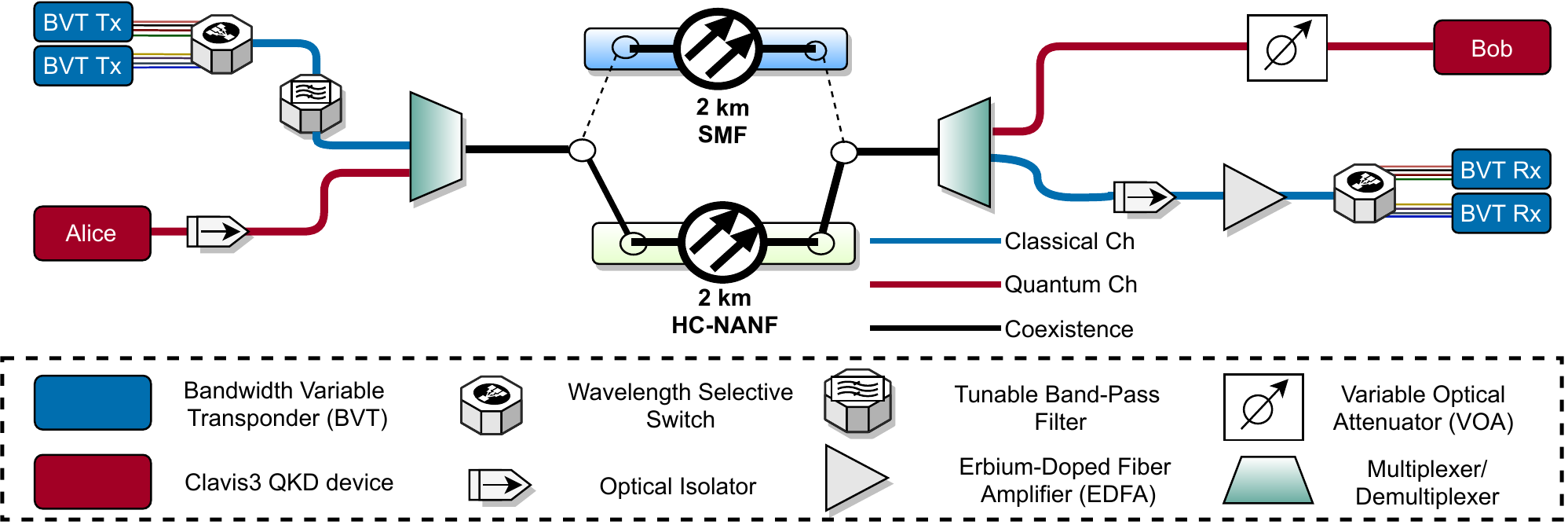}
    \caption{Experimental testbed design.}
    \label{fig:testbed}
\end{figure*}

\begin{figure}[H]
   \centering
        \includegraphics[width=\linewidth]{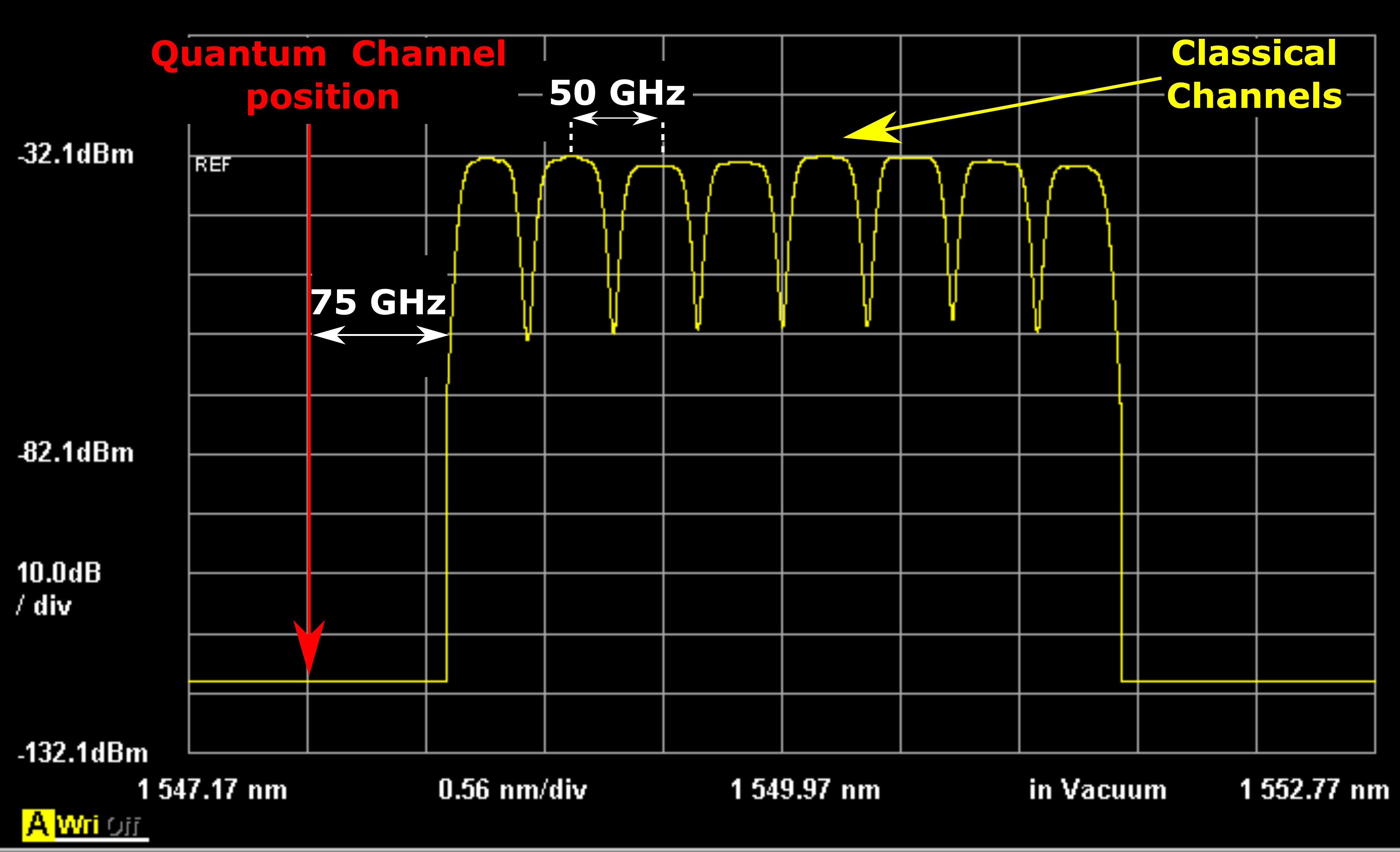}
    \caption{Coexistence spectrum  of  the   quantum  channel  (Red)  and  8$\times$200  Gpbs  classical  channels  (Yellow).}
    \label{fig:spectrum}
\end{figure}

\Cref{fig:testbed} shows the experimental system setup used to demonstrate the coexistence of eight classical channels with the quantum channel. The testbed facilitates the experimental evaluation of the nonlinear effects on the performance of the quantum channel created by the presence of classical channels spectrally close to the quantum channel using both HC-NANF and SMF. For the classical channels, two bandwidth variable transponder (BVT) switches (Facebook Voyager) are deployed to transmit eight coherent classical channels. The frequencies of the classical channels range from 193.60~THz to 193.25~THz with 50~GHz spacing between each channel and 75~GHz spacing between the quantum channel (193.70~THz) and the nearest classical channel as shown in \Cref{fig:spectrum}. Each classical channel provides a 200~Gbps, 16-QAM signal with a soft-decision forward error correction (SD-FEC) of 15\% to enable an error-free transmission. For the quantum channel, IDQuantique Clavis3 DV-QKD system \cite{CL3} is used which implements Coherent One Way (COW) protocol and has a fixed frequency for the quantum channel at 193.70~THz. As shown in \Cref{fig:testbed}, the eight coherent output ports of the two BVTs are multiplexed using a wavelength selective switch (WSS) with 6~dB of insertion loss for a total throughput of 1.6 Tbps. The WSS is used as a multiplexer and a band pass filter to couple the classical channels into a single fibre and provide a 30~dB isolation. The WSS combined output feeds the input of a tunable band pass filter (TBPF) with 60~dB of isolation to further suppress the noise generated by the classical channels. 
The classical channels and quantum channel are then coupled through a WDM multiplexer in a coexisting configuration and travels through a 2~km of SMF or HC-NANF to a WDM demultiplexer that passes the quantum channel (193.70~THz) while rejecting all other channels (classical channels). The rejected classical channels are forwarded towards an optical isolator with insertion loss of 2~dB to prevent the tunable laser used by the BVT Rx as a local oscillator from returning to the Bob-QKD unit and interfering with the QKD measurements. It also prevents the Amplified Spontaneous Emission (ASE) noise generated by the Erbium-Doped Fibre Amplifier (EDFA) which is used to amplify the classical signal travelling back to the Bob-QKD unit. The amplified classical channels are directed to the Voyager BVT Rx for coherent detection after the EDFA via a 1x8 optical splitter. 
Furthermore, an optical isolator is used after the Alice-QKD unit to prevent the noise generated from the classical channels returning to the Alice-QKD unit and interfering with the quantum signal. Moreover, the variable optical attenuator (VOA) is used before the Bob-QKD unit to adjust the losses of the quantum channel to the minimum operation level of 10~dB to prevent over saturating the single photon detector in the Bob-QKD unit. It also allows us to obtain a direct comparison of the quantum/classical coexistence between the SMF and HC-NANF in terms of losses. The loss of the quantum channel using the HC-NANF and SMF is $\approx$ 10.5~dB. Moreover, the 2~km HC-NANF used in this experiment has a loss of 1.3~dB/km \cite{bradley2018record} and a total loss of 7~dB when accounting for the  interconnection losses between the HCF and SMF. 

\section{Results}

\begin{figure*}[t]
\centering
\includegraphics[width=0.95\linewidth]{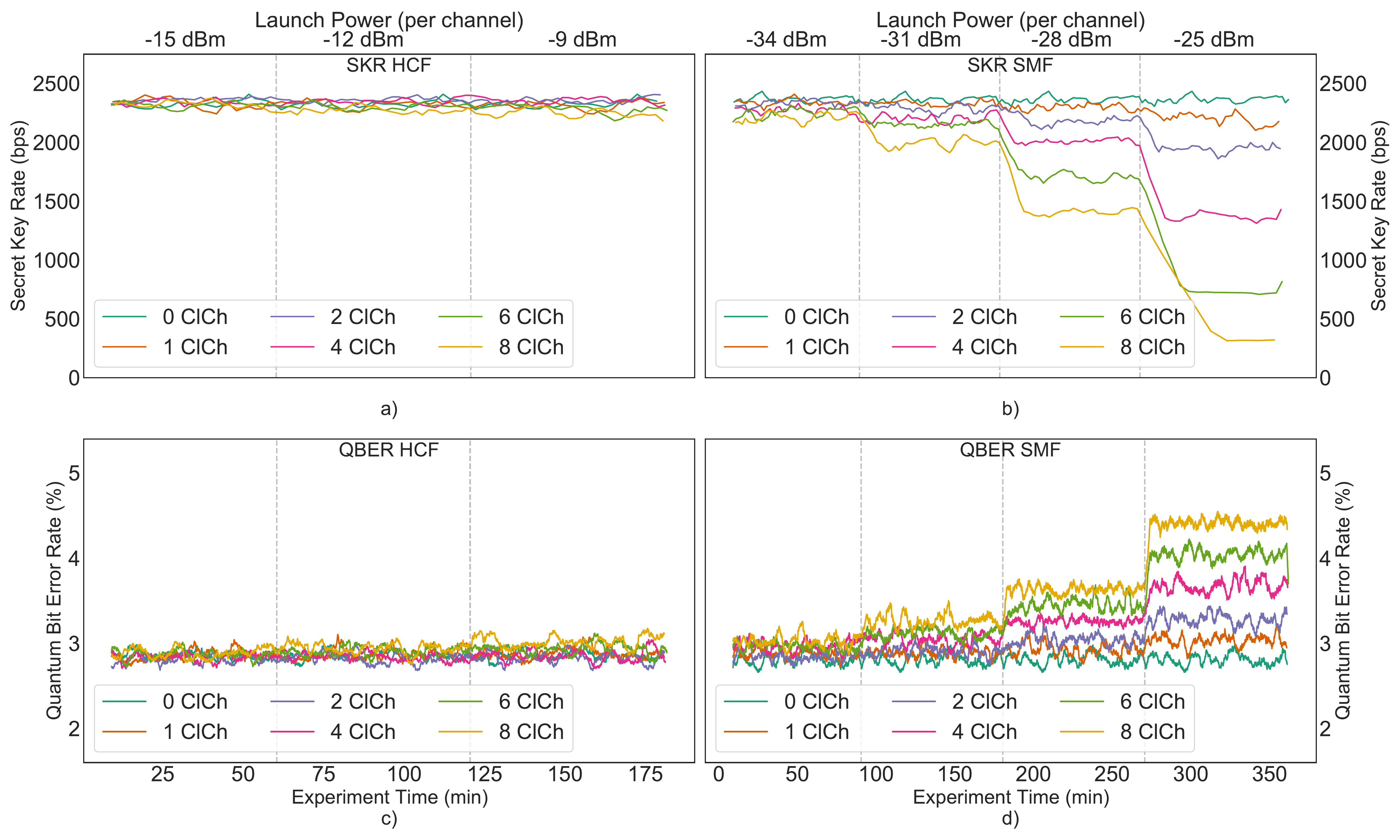}
\caption{a) Average SKR versus launch optical power using HC-NANF. b) Average SKR versus launch optical power using SMF. c)~Average QBER versus launch optical power using HC-NANF. d) Average QBER versus launch optical power using SMF.}

\label{fig:results}
\end{figure*}

\Cref{fig:results} shows the experimental evaluation of SKR and QBER of the quantum channel in the presence of a different number of classical channels at different launch powers for HC-NANF and SMF. \Cref{fig:results}a) and \Cref{fig:results}c) present the changes in the SKR and QBER of the quantum channel when coexisting with different number of classical channels at different launching power in a HC-NANF. As shown in \Cref{fig:results}a) and \Cref{fig:results}c), the SKR and QBER values without the presence of any classical channel (no coexistence) are similar to the values when coexisting 8 classical channels at a launch power of -9~dBm per channel which is equivalent to the highest recorded coexistence power of 0~dBm. This is due to the ultra-low optical nonlinearity in HC-NANF. Moreover, \Cref{fig:results}b) and \Cref{fig:results}d) show the SKR and QBER when coexisting quantum and classical channels in SMF. As shown in \Cref{fig:results}b) and \Cref{fig:results}d) the SKR values drops from $\approx$ 2400 bps to $\approx$ 304 bps while the QBER values increase from $\approx$ 3\% to $\approx$ 4.5\% when coexisting 8 classical channels at a launch power of -25~dBm per channel which is equivalent to a coexistence power of -16~dBm. This significant 87.5\% drop in the SKR and 50\% rise in the QBER is due to high optical nonlinearity in SMF causing a high noise leakage to the Bob-QKD unit. Moreover, increasing the coexistence power by 1~dBm would cause the SKR to plummet to zero bps. Although the losses of the quantum channel are similar when using both SMF and HC-NANF and the coexistence power (0~dBm) in the HC-NANF is 40~times higher than coexistence power (-16~dBm) in SMF, the ultra-low nonlinear effects of the HC-NANF due to its hollow core preserve the SKR and QBER of the quantum channel. This proves the suitability of HC-NANF as an excellent transmission medium for coexistence of quantum and classical channels.

\section{Conclusions}


The coexistence of a DV-QKD channel and 8~$\times$~200~Gbps classical channels was successfully demonstrated over a 2~km long HC-NANF for a record-high transmission of 1.6 Tbps. The SKR was preserved without any noticeable changes when coexisting the quantum channel with eight classical channels at 0~dBm coexistence power compared to a significant drop of 87.5\% when using SMF at -16~dBm coexistence power which is 40~times lower than the power used in HC-NANF. This significant difference in the QKD performance proves the advantage of using HC-NANF to provide a seamless coexistence of quantum and classical channels with a minimal effect on the quantum channel performance regardless of the number and power of the classical channels.


\section{Acknowledgements}
\small
This work was funded by EU funded project UNIQORN (820474) and the EPSRC Airguide Photonics Partnership Fund Award (ref: 517129) (EP/P030181/1). Part of the research leading to this work has been supported by the Quantum Communication Hub funded by the EPSRC grant ref. EP/T001011/1 and the ERC LightPipe project (grant n 682724).


\clearpage
\printbibliography

\vspace{-4mm}

\end{document}